\documentclass[fleqn,usenatbib]{mnras}

\usepackage{newtxtext,newtxmath}

\usepackage[T1]{fontenc}

\DeclareRobustCommand{\VAN}[3]{#2}
\let\VANthebibliography\thebibliography
\def\thebibliography{\DeclareRobustCommand{\VAN}[3]{##3}\VANthebibliography}

\usepackage{graphicx}	
\usepackage{amsmath}	
\usepackage{amssymb}	
\usepackage{subfig}
\usepackage{comment, bm}
\graphicspath{{./}{figures/}}

\usepackage{censor}
\usepackage{ulem}

\usepackage{filecontents}
\usepackage{ftnxtra}
\usepackage{calc}
\newlength\nextcharwidth
\makeatletter
\renewcommand\@cenword[1]{%
  \setlength{\nextcharwidth}{\widthof{#1}}%
  \censorrule{\nextcharwidth}%
  \kern -\nextcharwidth%
  #1}
\makeatother

\title[ArPLS-ST]{Radio Frequency Interference Mitigation based on the ArPLS and SumThreshold Method}

\author[Zeng et al.]{
Qingguo Zeng,$^{1}$
Xue Chen,$^{2,3}$
Xiangru Li,$^{4}$\thanks{E-mail: xiangru.li@gmail.com (X.R. Li)}
J. L. Han,$^{2,3,5}$
Chen Wang,$^{2,3,5}$
\newauthor{
D. J. Zhou,$^{2,3}$
Tao Wang$^{2,3}$}
\\
$^{1}$School of Mathematical Sciences, South China Normal University, No. 55 West of Yat-sen Avenue, Guangzhou 510631, China\\
$^{2}$National Astronomical Observatories, Chinese Academy of Sciences,
Jia20 Datun Road, Beijing 100012, China\\
$^{3}$Astronomy School, the University of Chinese Academy of Sciences,
No.19(A) Yuquan Road, Shijingshan District, Beijing 100049, China\\
$^{4}$School of Computer Science, South China Normal University, No. 55 West of Yat-sen Avenue, Guangzhou 510631, China\\
$^{5}$The FAST key laboratory, Chinese Academy of Sciences,
Jia20 Datun Road, Beijing 100012, China
}

\date{Accepted 2020 August 17. Received 2020 August 12; in original form 2020 March 18}

\pubyear{2020}

\begin{document}
\label{firstpage}
\pagerange{\pageref{firstpage}--\pageref{lastpage}}
\maketitle

\begin{abstract}
As radio telescopes become sensitive, radio frequency interference (RFI) is more and more serious for interesting signals of radio astronomy. There exist demands for developing an automatic, accurate and efficient RFI mitigation method. Therefore, this work investigated the RFI detection algorithm. Firstly, we introduced an Asymmetrically Reweighted Penalized Least Squares (ArPLS) method to estimate baseline more accurately. After removing the estimated baseline, several novel strategies were proposed based on the SumThreshold algorithm for detecting different types of RFI. The threshold parameter in the SumThreshold can be determined automatically and adaptively. The adaptiveness is essential for reducing human interventions and the online RFI processing pipeline. Applications to FAST (Five-hundred-meter Aperture Spherical Telescope) data show that the proposed scheme based on the ArPLS and SumThreshold is superior to some typically available methods for RFI detection with respect to efficiency and performance.
\end{abstract}




\section{Introduction} \label{sec:intro}

In radio astronomy, radio frequency interference (RFI) becomes more and more serious for radio observational facilities. The RFI always influences the searching and analysis of the interesting astronomical objects.
Mitigating the RFI becomes an essential procedure in pulsar survey data processing. Five-hundred-meter Aperture Spherical radio Telescope (FAST) is an extremely sensitive radio telescope formally in operation in January 2020. It is necessary to find out an effective and precise RFI mitigation method for FAST data processing.

Available RFI mitigation methods can be divided into three categories based on their principles \citep{akeret2017radio}. The first category is of linear methods, such as, Singular Vector Decomposition \citep[SVD; ][]{offringa2010post}, Principle Component Analysis (PCA) and their variants \citep[e.g. ][]{zhao2013windsat}. In practice, these methods are not suitable for dealing with frequency-varying RFI \citep{offringa2010post}. The widespread use of radio sources in human life causes the diversity of the RFI. The diverse contamination of RFI makes the RFI difficult to be modeled by these linear methods. The second category is of machine learning schemes. The machine learning algorithms can automatically learn the discriminant features between RFI and non-RFI \citep{mosiane2016radio,akeret2017radio, kerrigan2019optimizing}. One typical limitation of this kind methods is that they need a set of observations with labels which are time-consuming to be obtained. The last category is thresholding method widely used in the available RFI mitigation pipelines due to its simpleness and effectiveness. One typical thresholding method is the simple thresholding \citep{schoemaker2015removing}, which flags a pixel as RFI in case of its intensity larger than a preset parameter (called threshold). The superiorities of this method are its simplicity and high efficiency. However, this method is sensitive to noise due to its dependencies on single pixel comparison. To overcome this limitation, \cite{offringa2010post} introduced an RFI detection algorithm, SumThreshold, based on computing the combinatorial effects of some adjacent pixels. The SumThreshold method has been wrapped in the RFI detection pipeline for the Low Frequency Array (LOFAR), e-MERLIN \citep{peck2013serpent}, Bleien Radio Observatory \citep{akeret2017hide}, etc.

In the thresholding RFI detection methods, a fundamental assumption is that the intensities of the data should be constant in the absence of interferences \citep{winkel2007rfi}. However, almost all of the astronomical data do not fit this assumption due to the presence of the inconsistency of receiver response and background information. This kind inconsistency has some negative impacts on the RFI detection and can be approximately described using a smooth surface \citep[referred to as baseline, ][]{winkel2007rfi}. The baseline should be accurately estimated and removed from data. To do this, \cite{winkel2007rfi} proposes a scheme to describe the baseline using a two-dimensional, low order polynomial; \cite{offringa2010post} proposes a baseline fitting scheme based on a sliding window and some weighted Gaussian filters. However, it is shown that the accuracy of these baseline estimations can be affected by a broad-band RFI.

Therefore, we proposed a baseline fitting method based on an Asymmetrically Reweighted Penalized Least Squares algorithm \cite[ArPLS; ][]{baek2015baseline}. The penalized constraint in this method makes the baseline fitting more robust and accurate than traditional methods by mitigating the negative influences from instrumental response. The baseline is estimated from a time-integral curve/Spectral Energy Distribution (SED) curve (a 1-dimensional vector); while the traditional method is done in a time-frequency image \citep{offringa2012algorithms}. Therefore, this ArPLS-based method is more efficient.

For flagging the RFI, we gave several strategies based on the SumThreshold. These strategies not only can detect the traditional band RFI more efficiently, but also can more accurately detect blob RFI, a short and small-bandwidth interference typically covering nearly 100 microseconds and a bandwidth of less than one MHz.

This article is organized as follows: The experimental FAST data and their characteristics are described in section \ref{sec:data}. In section \ref{sec:method}, we presented the proposed baseline fitting method and the strategies to detect the RFI. The application of the proposed scheme to FAST data and some discussions are carried out in Section \ref{sec:Results_Discussion}.

\section{Experimental data and their characteristics}
\label{sec:data}

The proposed RFI mitigation scheme is tested on FAST observations. These data are sampled at a time-resolution $4.9152\times10^{-5}$ seconds on 4096 frequency channels. The size of each time-frequency image is 4096$\times$1024 pixels, where 1024 is the number of sampling points per frequency channel within one sub-integration.

The data set consists of 100 time-frequency images (sub-integrations) which can be taken for any beam in different sky-areas, and at different observation time by the 19-beam receiver. The diversity of RFI and baseline guarantee the objectiveness of the performance evaluation on the proposed scheme. To design a RFI mitigation scheme as efficient and accurate as possible, it is necessary to investigate the characteristics of the RFI on the FAST data.

\begin{figure*}
  \centering%
    \includegraphics[width=0.9\linewidth]{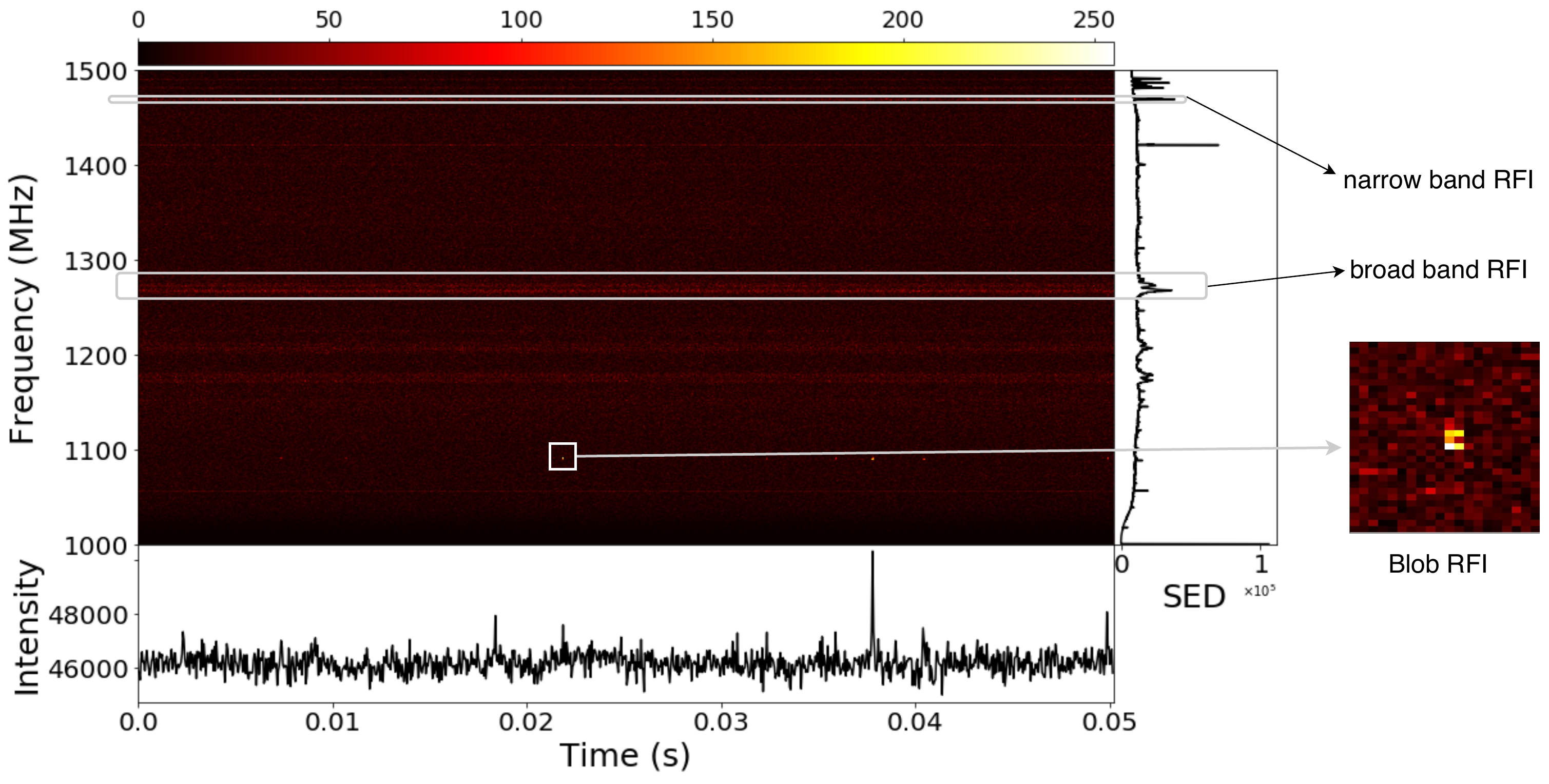}
\caption{A time-frequency image of FAST data for 0.05 seconds in the frequency range from 1000 to 1500 MHz with a SED curve in the right, a frequency-integral curve in the lower panel, and a zoomed views of a RFI region as the main panel. A narrow band RFI and a broad band RFI can be identified from the peaks on the SED curve, the blob RFI contaminates only a small ratio of pixels and cannot be identified based on the SED curve. The colorbar of the image is shown on the top of the time-frequency image.}
\label{fig:fastdata}
\end{figure*}

The pulsar search observations of FAST take a wavelength range from 1000 to 1500 MHz and frequency resolution 122.07 KHz \citep{jiang2020fundamental}. In FAST observations, there are mainly two types of RFI: band RFI and blob RFI (Figure \ref{fig:fastdata}). The band RFI is likely to be generated by TV broadcasts, mobile communication and radar. The blob RFI is a short, small-bandwidth signal from unknown sources.
Suppose $s(t, f)$ represents the input `Data' in Figure \ref{fig:flowchart}, where $t$ represents time and $f$ the frequency. The SED is computed by aggregating the energies along the time axis $\texttt{SED}(f) =\sum\limits_t{s(t, f)}/n_{t}$ (Figure \ref{fig:fastdata}), where $n_t$ is the number of pixels per frequency channel in one sub-integration. The band RFI occupies one or several frequency channels with a time-duration of almost the whole sub-integration; while the blob RFI just contaminates several pixels. Figure \ref{fig:fastdata} shows one typical FAST observation and the corresponding SED curve. In practice, there may be more than one peak on one SED segment contaminated by one frequency-varying band RFI \citep{jiang2020fundamental}.

\section{The proposed scheme}\label{sec:method}
\begin{figure*}
	\centering
	\includegraphics[width=0.9\textwidth]{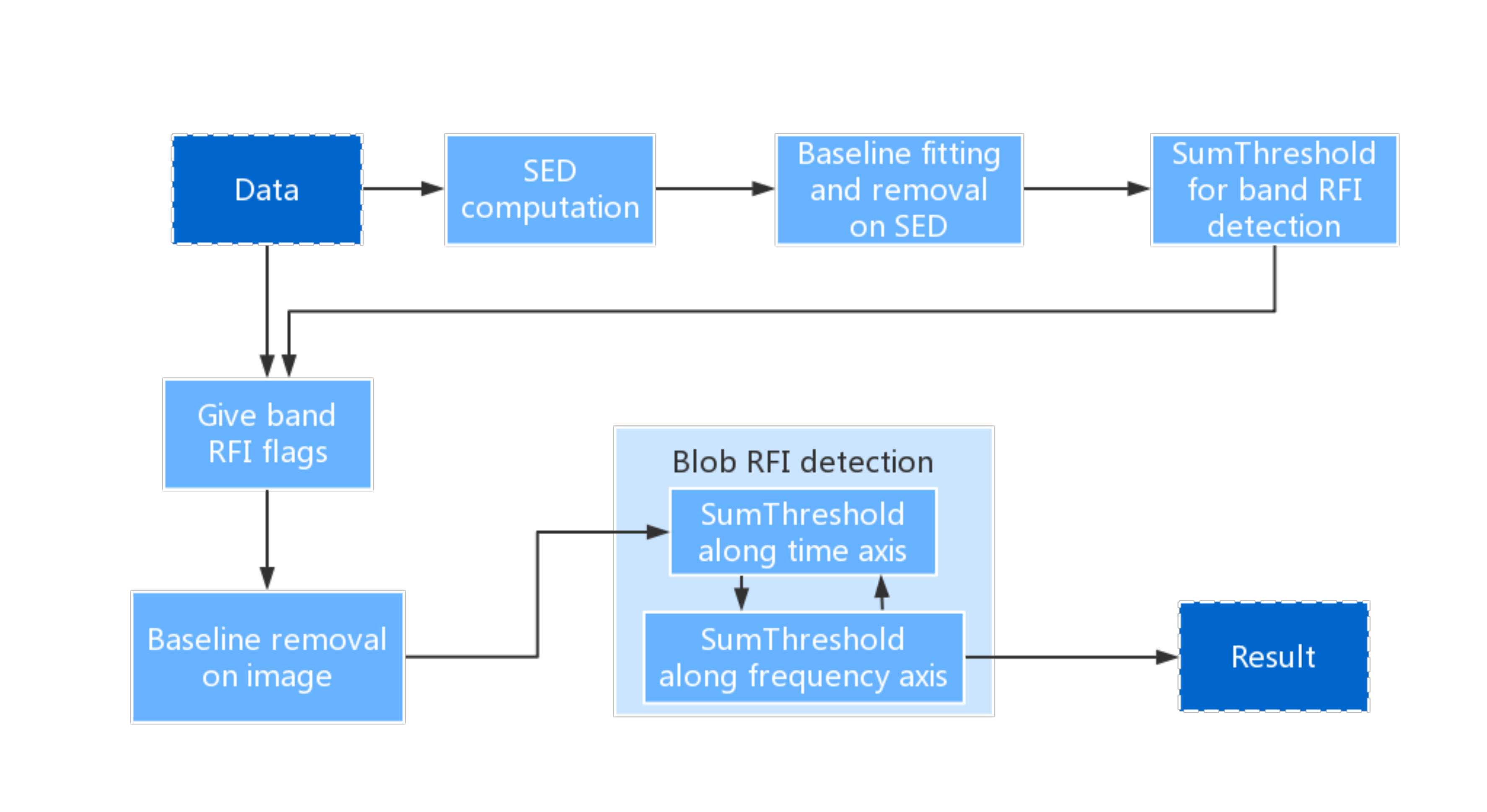}
	\caption{A flowchart of the ArPLS-ST scheme. The `Data' is an observation of time-frequency image $s(t, f)$. The baseline-fitting and removal are designed to reduce the negative effects on RFI detections from the inconsistency of receiver response and background information.}
	\label{fig:flowchart}
\end{figure*}

Based on the characteristics of the RFI in the FAST data, we propose a novel RFI mitigation scheme. This RFI mitigation scheme is designed based on the two main parts as being the asymmetrically reweighted penalized least squares (ArPLS) and SumThreshold (ST). For convenience, this scheme is referred to as ArPLS-ST. A flowchart of the ArPLS-ST is presented in Figure \ref{fig:flowchart}. The core procedures are `Baseline fitting and removal on SED', `SumThreshold for band RFI detection', `Baseline removal on a time-frequency image' and `Blob RFI detection'. For fitting the baseline, we introduce the ArPLS method. In the procedures `Band RFI mitigation' and `Blob RFI mitigation', several novel strategies based on the ST algorithm  are applied to detect different types of RFI. Besides, the threshold in ArPLS-ST can be automatically determined by a generalized PauTa criterion \citep{shen2017two}. This automatic parameter setting reduces manual interventions and makes the scheme suitable for an automatic processing pipeline.

\subsection{Baseline fitting and removal}\label{sec:method:baselinefitting}

Instead of estimating the baseline in the time-frequency image, we proposed an estimation from the SED curve using the ArPLS algorithm.

\subsubsection{The ArPLS for fitting the baseline on the SED curve}\label{sec:method:baselinefitting:ArPLS}

A suitable baseline estimation should satisfy two requirements: fitness and smoothness. Let $\mathbf{y} \in R^D$ denote the data under being processed, $\mathbf{z} \in R^D$ the estimated baseline of $\mathbf{y}$, where $D$ is a positive integer and denotes the number of sampling points along the frequency axis. In this work, $\mathbf{y}$ represents the SED curve of an observation (a sub-integration from FAST). The constraint `fitness' ensures that the estimated baseline $\mathbf{z}$ precisely describes the information of the original signal $\mathbf{y}$ within interference-free regions; while the latter constraint, `smoothness', ensures the estimated baseline is not influenced by the RFI. Consequently, the optimal estimation of $\mathbf{z}$ can be obtained by minimizing the following weighted penalized least squares function \citep{eilers2003a, cobas2006new, zhang2010intelligent, baek2015baseline},
\begin{equation}
    S(\mathbf{z}) = (\mathbf{y} - \mathbf{z})^\top\mathbf{W}(\mathbf{y} - \mathbf{z})+\lambda \mathbf{z}^\top\mathbf{M}^\top\mathbf{M}\mathbf{z},
    \label{equ:S2}
\end{equation}
where $\mathbf{W}$ is a diagonal matrix with its diagonal element $\mathbf{w_i} \geq 0$ representing the weight corresponding to the square difference $(\mathbf{y}_i - \mathbf{z}_i)^2$, $i= 1, \cdots, D$; $\mathbf{M}$ is a $D\times D$ matrix. Actually, the $\mathbf{M}$ is a second order difference matrix which is considered as a natural way to express the roughness in mathematics \citep{ramsay2007applied}. Besides, $\lambda$ is a preset coefficient that controls the balance between fitness and smoothness. Ideally, $w_i$ should be set to a value almost zero for the pixels in the peak regions contaminated by RFI and nearly one for the pixels outside these regions. Unfortunately, these peak regions remain unknown for a given observation and it is difficult and time-consuming to locate them in application \citep{andreev2003universal, jirasek2004accuracy}. \cite{baek2015baseline} proposed an iteratively weighting procedure to obtain the optimal estimation of $\mathbf{z}$ and $\mathbf{W} $ without peak searching. This iteratively weighting procedure is referred to as ArPLS algorithm.


\subsubsection{Baseline fitting and removal on SED curve}\label{sec:method:baselinefitting:SED}
The SED curve can be divided into three parts according to their RFI-contamination characteristics: interference-free regions, narrow band RFI regions and broad band RFI regions. In the interference-free regions, the SED curve is smooth. On the other hand, the band RFI causes some dramatic fluctuations (Figure \ref{fig:fastdata}). There are sometimes multiple peaks within one protuberance in the regions contaminated with some broad band RFI, which inevitably induce some difficulties in baseline fitting.

\begin{figure*}
\centering
\includegraphics[width=1\linewidth]{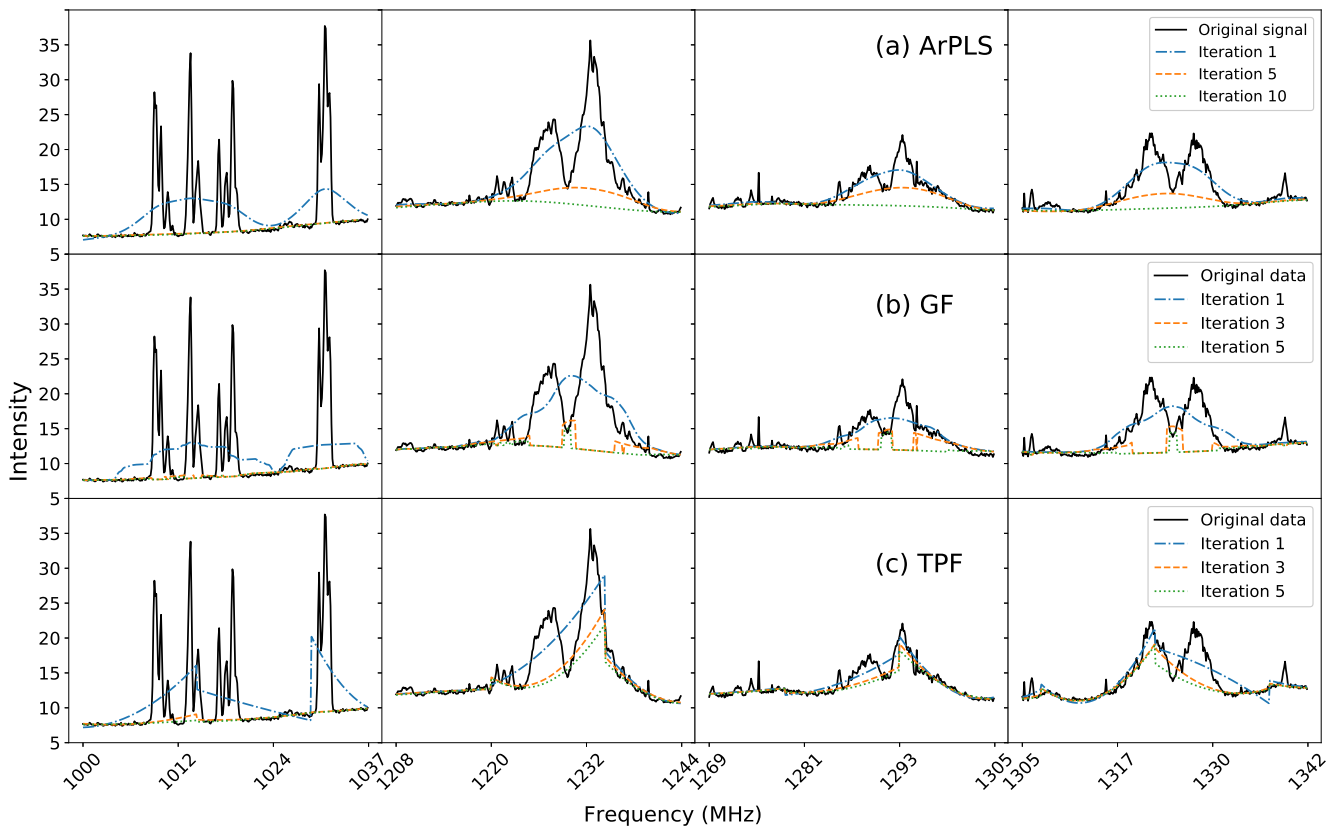}
\caption{Examples of baseline fitting using three baseline fitting methods (ArPLS, TPE, and GF) to the observation data in the frequency ranges 1000-1037 MHz, 1208-1244 MHz, 1269-1305 MHz and 1305-1342 MHz. In these examples, there exist several narrow band RFI regions and broad band RFI regions. The convergence characteristics of the iterative process are shown for several iteration steps.}
\label{fig:local}
\end{figure*}

It is shown that ArPLS can quickly converge in the interference-free regions and narrow band RFI regions (Figure \ref{fig:local} a). In the broad band RFI regions, although the ArPLS converges relatively slowly, experiments show that it is still capable of giving a reasonable estimation for the baseline after several more iterations (Figure \ref{fig:local} a). To our knowledge, typical baseline fitting methods used in pulsar data processing are the tile-based polynomial fitting \citep[TPF; ][]{winkel2007rfi} and the weighted Gaussian filter \citep[GF; ][]{offringa2010post}. It is shown that both the TPF and GF work well in the interference-free regions and narrow band RFI regions (Figures \ref{fig:local} b and c). However, the baselines fitted by them tend to be raised up by the peaks within broad RFI regions. Furthermore, the TPF method performs poorly near the edges of each tile due to the boundary effects of the polynomial fitting, especially in case the edge is in the peak regions.

The most significant difference between these two methods and the proposed ArPLS is that the ArPLS can fit the baseline directly by tolerating the RFI in the data, while the TPF and GF couple the RFI removing and baseline fitting due to their sensitivity to the RFI. In case the sharp peaks are marked as candidate RFI regions by the TPF and GF, the pixels within these regions are discarded. However, this discarding makes these methods difficult to accurately estimate a smooth baseline and judge whether the regions between the marked regions are contaminated with any relatively weak RFI or not. Therefore, the TPF and GF often fail to detect some relatively weak RFI on the regions between two strong peaks in the broad band RFI regions (Figure \ref{fig:local} b and c).

Actually, the smoothness of the baseline estimated by the proposed scheme is controlled by the second term of equation (\ref{equ:S2}). This constraint is implemented using a second order difference and adaptive to the radio observations in the iterative estimation procedures. In the TPF and GF, however, the smoothness is constrained by the order of the polynomial and the scale parameter respectively, which are preset based on human experiences. Therefore, the proposed ArPLS method is more robust than the TPF and GF.

Although the ArPLS is an iterative algorithm and there are some equations to be solved in each iteration, experiments show that it is still fast enough because the system is sparse and almost all calculations can be implemented in a vectorized style. The running time per execution of three baseline fitting methods is presented in Table \ref{tab:time}. The ArPLS is the fastest in these three methods, and the other two methods need to be executed several times for every time-frequency image in the FAST application.

\begin{table}
	\centering
	\caption{The average execution time of the baseline fitting methods for one FAST time-frequency image. They are computed by running every method for 10 times.}
	\label{tab:time}
	\begin{tabular}{ll} 
\hline
Methods      & Execution time\\
\hline
ArPLS                          & $27.4  \pm 0.4 ms$ \\
Weighted Gaussian filter (GF) & $32.7\pm 0.4ms$ \\
Tiled-based polynomial fitting (TPF) & $38.3  \pm 0.6ms$ \\
\hline
	\end{tabular}
\end{table}

\begin{figure*}
\centering
\includegraphics[width=1\linewidth]{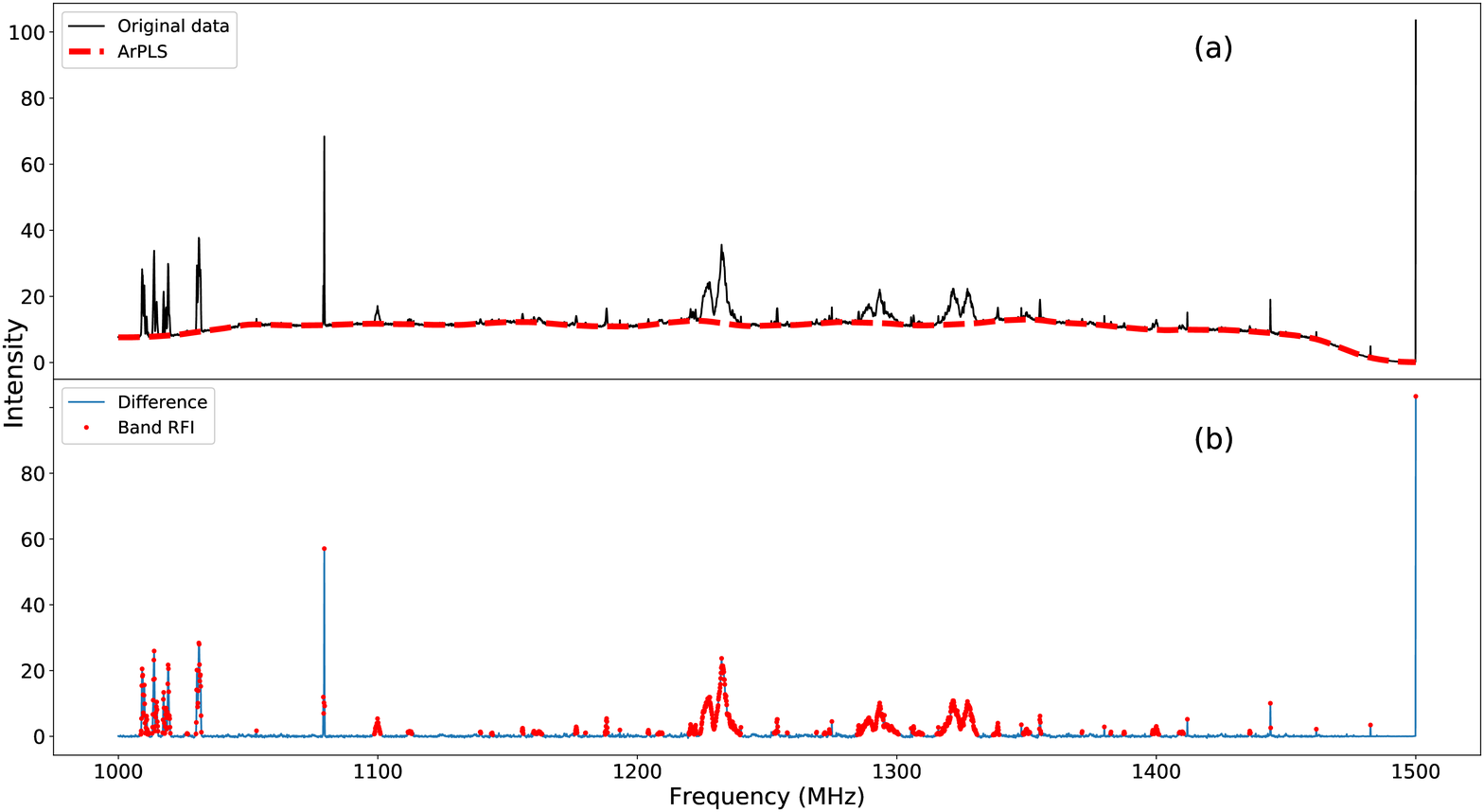}
\caption{Results of baseline fitting and removal using the ArPLS for the time-frequency image in Figure \ref{fig:fastdata}. (a) An original SED curve (solid line) and the baseline (dashed line) fitted by the ArPLS. (b) The SED curve after removing the estimated baseline (solid line), and the frequency channels corresponding with the detected band RFI (marked as points).}
\label{fig:global}
\end{figure*}

\subsubsection{Baseline removal on the time-frequency image}\label{sec:method:baselinefitting:image}

\begin{figure*}
	\centering
	\includegraphics[width=0.8\textwidth]{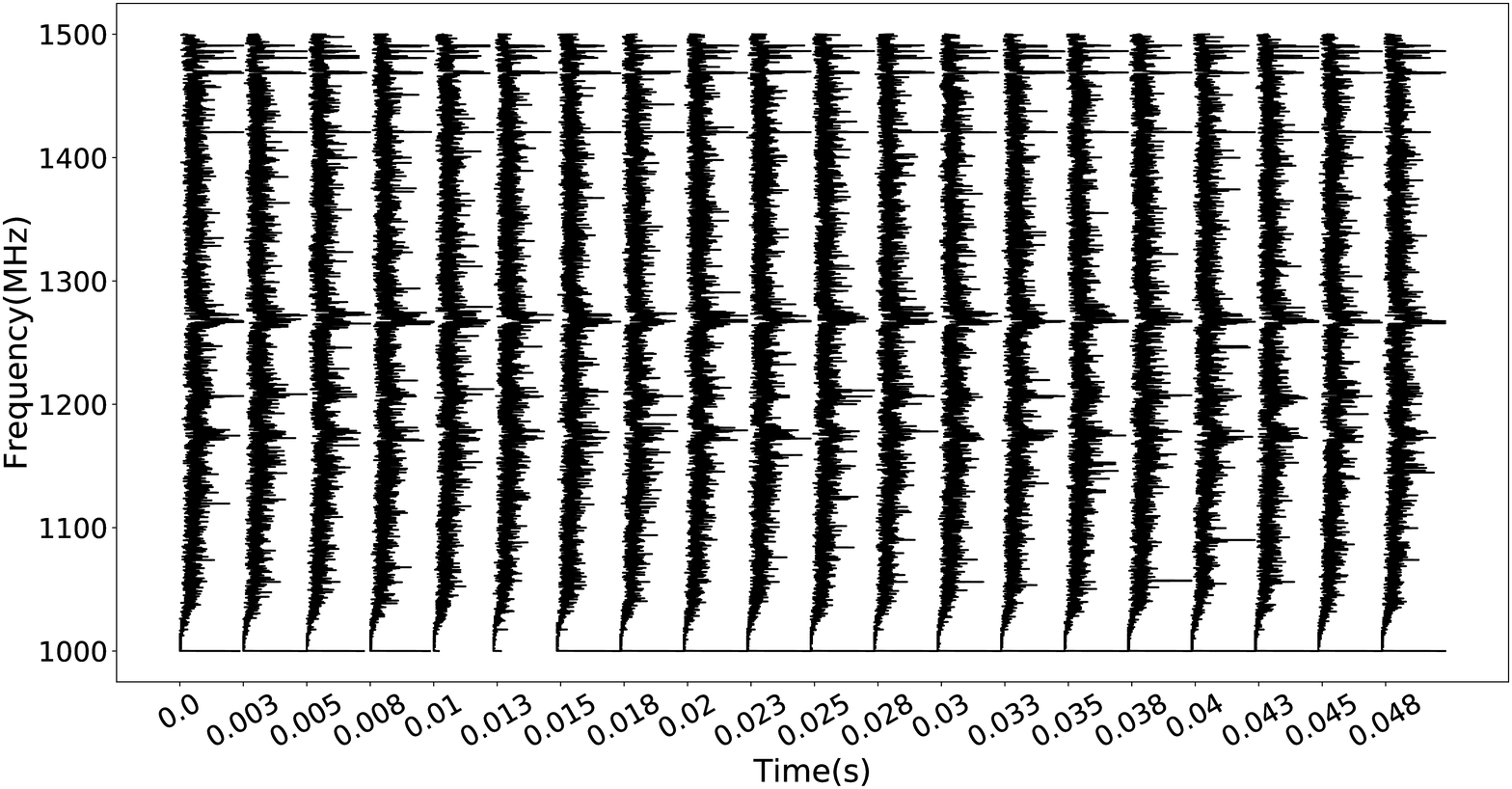}
	\caption{The stableness of SED with time for a set of sub-integration from FAST observations. This stableness indicates that the baseline of a time-frequency observation from FAST can be estimated using the SED curve.}
	\label{fig:bl_timestep}
\end{figure*}

To detect blob RFI, the baseline removal should be performed on the time-frequency image. However, it is time-consuming to estimate the baseline directly in a time-frequency joint space due to a large number of observation pixels. Fortunately, it is shown that the general tendency of the spectrum in a sub-integration from FAST observation is stable on time (Figure \ref{fig:bl_timestep}). Therefore, the baseline of each spectrum can be approximated by a shared curve theoretically, and this work used the baseline estimated from the SED curve as the shared curve.

\begin{figure*}
\centering
\subfloat[An original time-frequency image.]{
	\includegraphics[width=0.45\linewidth]{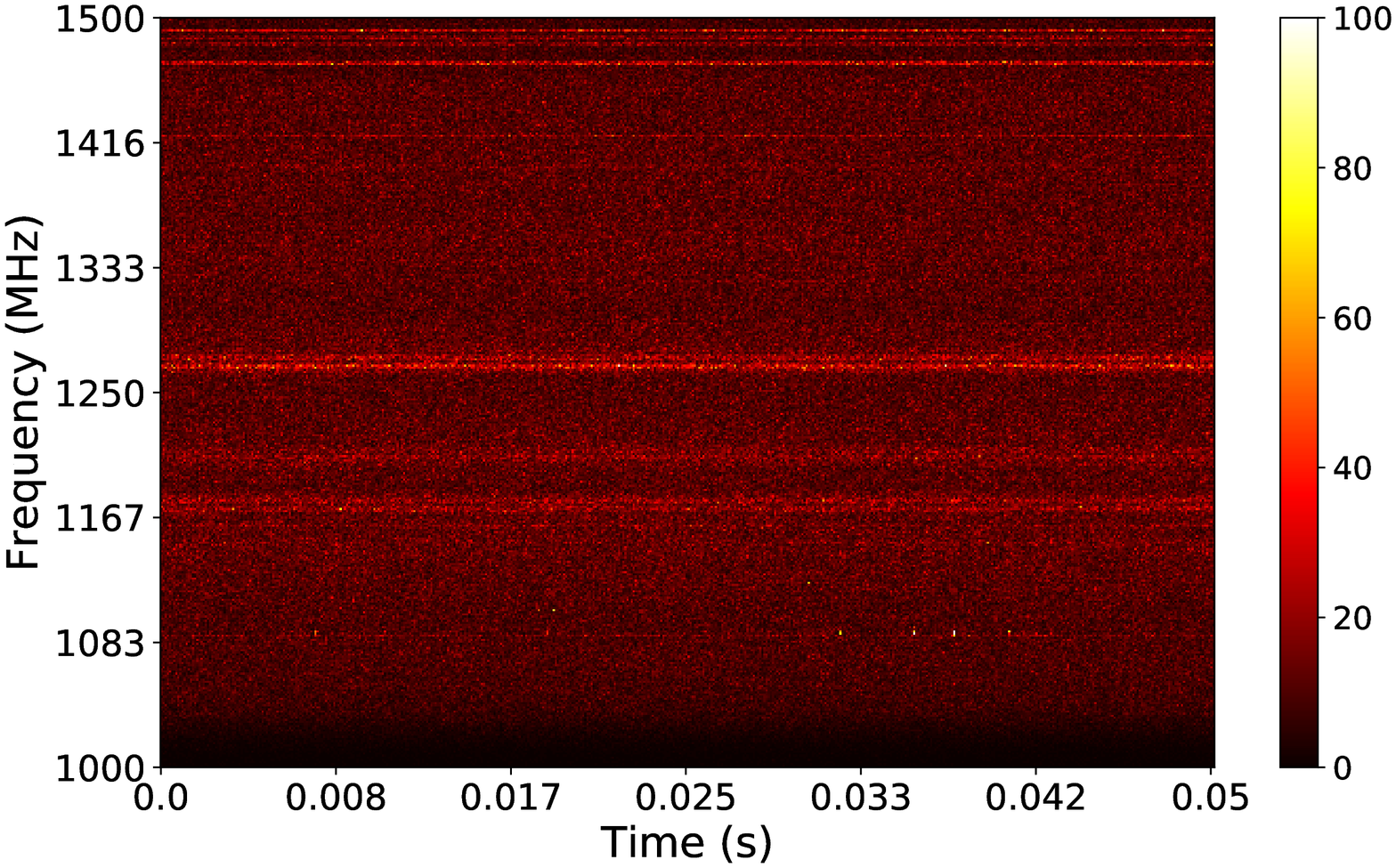}
	\label{fig:tf_before_bl}
	}
\subfloat[The result after baseline removal.]{
	\includegraphics[width=0.45\linewidth]{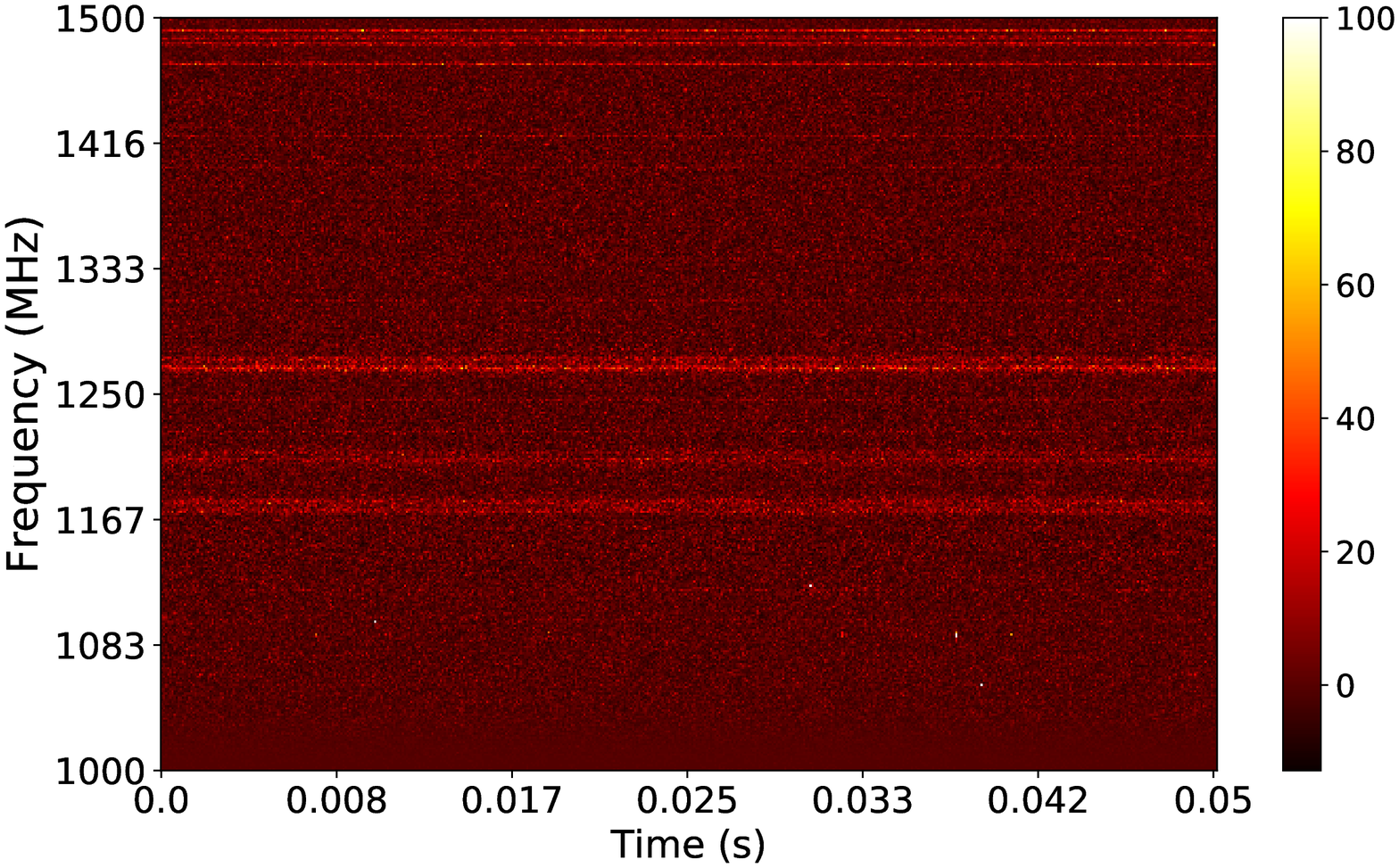}
	\label{fig:tf_after_bl}
	}
\caption{Comparison between a time-frequency image and the result after baseline removal.}
\label{fig:bl_tic}
\end{figure*}

The Figure \ref{fig:bl_tic} presents the result of baseline removal on a time-frequency image. Compared with the original time-frequency image, the background inconsistency of the processed image is removed excellently and the area with low contrast in the original image becomes easier to be distinguished. Therefore, the blob RFI can be identified more accurately using the thresholding algorithms after baseline removal. Meanwhile, this scheme saves computing resources and time by avoiding estimating the baseline for a time-frequency image from scratch.


\subsection{RFI detection based on the SumThreshold}\label{sec:method:RFI}

After baseline removal, the pixel intensity should be almost constant in the interference-free regions while peaks caused by the RFI still remain and are even more prominent (Figures \ref{fig:global}b and \ref{fig:tf_after_bl}). Therefore, we can accurately detect RFI using the SumThreshold method.

\begin{figure}
    \centering
    \includegraphics[width=0.9\linewidth]{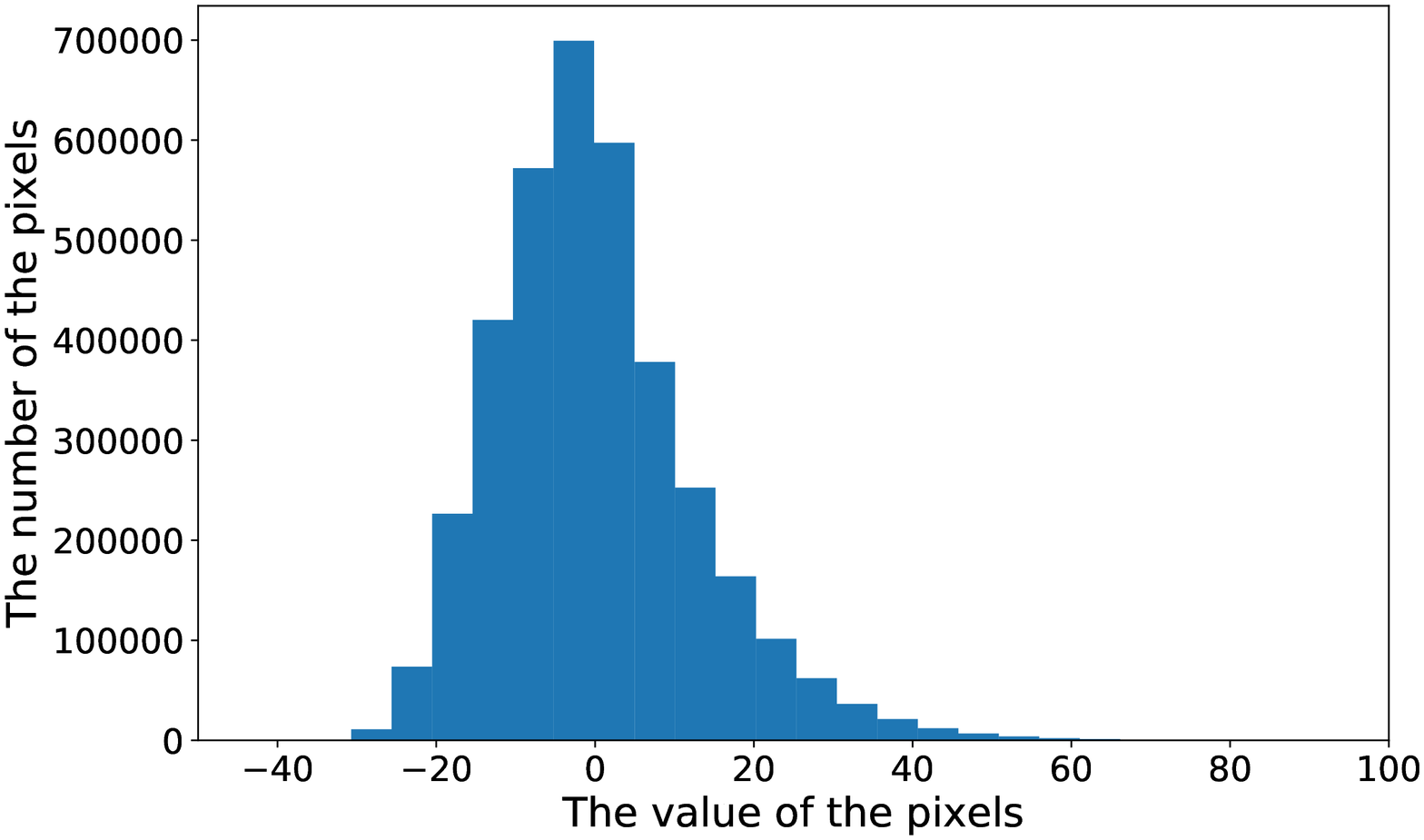}
    \caption{A histogram of pixel values  after baseline removal. This histogram can be approximated using a normal distribution. Note that the figure is truncated on value 100 in the x-axis to show the most of the pixels.}
    \label{fig:hist}
\end{figure}

The input to the SumThreshold is a one-dimensional vector which is the SED curve in the band RFI detection. For blob RFI detection, the input to the SumThreshold is a row or a column of a matrix representing a time-frequency observation after baseline removal. The SumThreshold is an iterative algorithm. In each iteration, four computational steps are carried out: calculating threshold, value replacement, summation, and RFI detection \& flagging. The fundamentals of the SumThreshold can be found in \cite{offringa2010post}. The $K \sigma$ criterion (a variant of PauTa criterion) is applied to adaptively determine the value of the threshold. Specifically, RFI is detected by checking whether a pixel deviates from the mean more than $K$ times of the standard deviation. The $K \sigma$ criterion determines the threshold based on the pixel-value distribution of the input to the SumThreshold (Figure \ref{fig:hist}). Some excellent investigations on the estimation of standard deviation can be found in \cite{Journal:Fridman:AJ2008}.

\subsubsection{Band RFI detection}\label{sec:method:RFI:BandRFI}

The SumThreshold is subsequently applied to the SED curve for detecting the band RFI and to a time-frequency image for detecting the blob RFI. Actually, the order, in which the SumThreshold is used for detecting the band RFI or blob RFI, does not have any influence on the detection results. However, we can detect RFI more efficiently by using the SumThreshold-based scheme in the order of band RFI first, especially in case of the existence of too many band RFI. After that, in detecting blob RFI, the SumThreshold does not need to be performed on the regions where band RFI is detected.

\begin{figure*}
\centering
	\subfloat[An original time-frequency image.]{
	\includegraphics[width=0.45\linewidth]{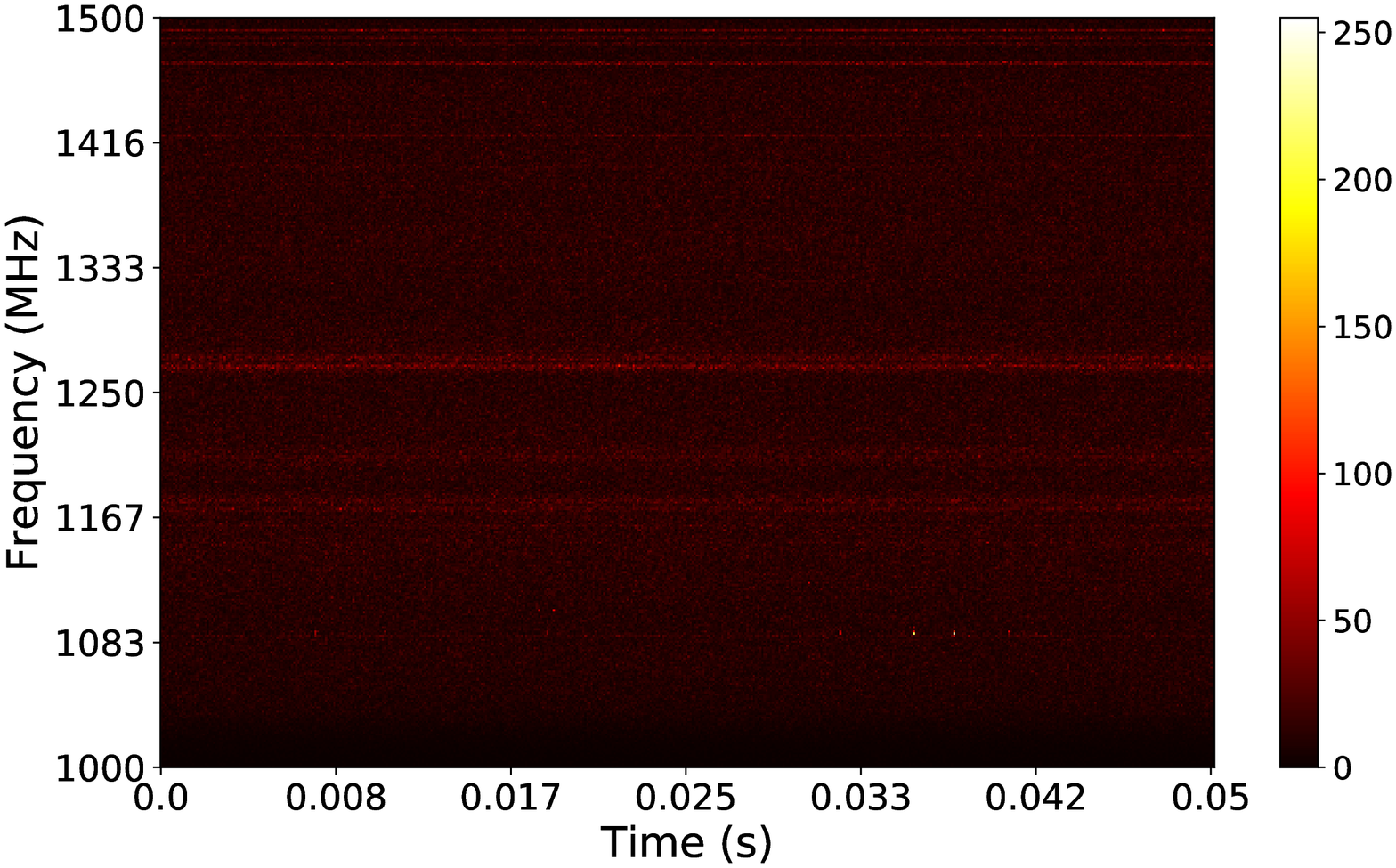}
	\label{fig:tf_ori}
	}
	\subfloat[The result after removing the band RFI using ArPLS-ST.]{
	\includegraphics[width=0.45\linewidth]{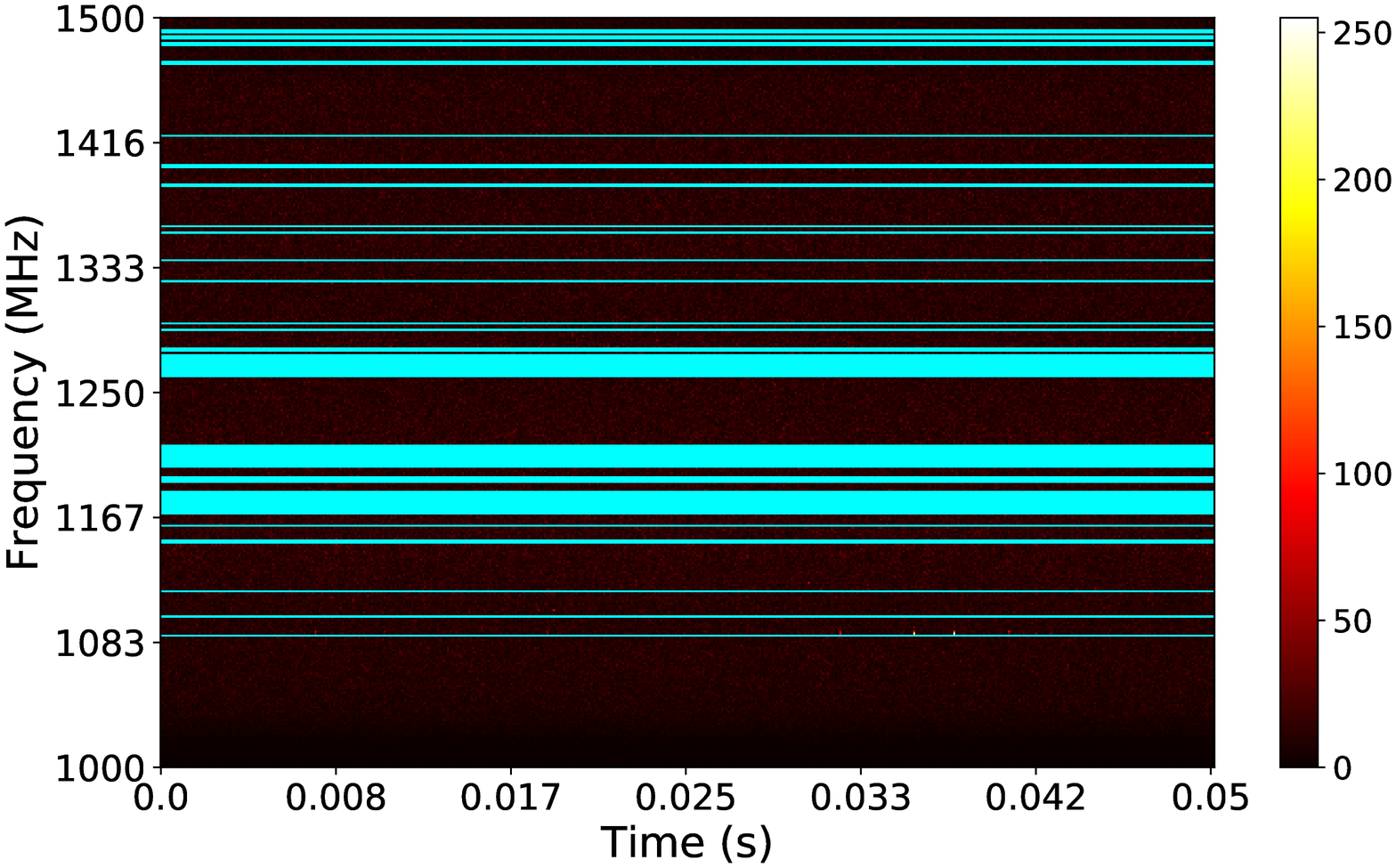}
	\label{fig:tf_band}
	}
	\vfill
	\subfloat[The result after removing the band RFI and blob RFI using ArPLS-ST.]{
	\includegraphics[width=0.45\linewidth]{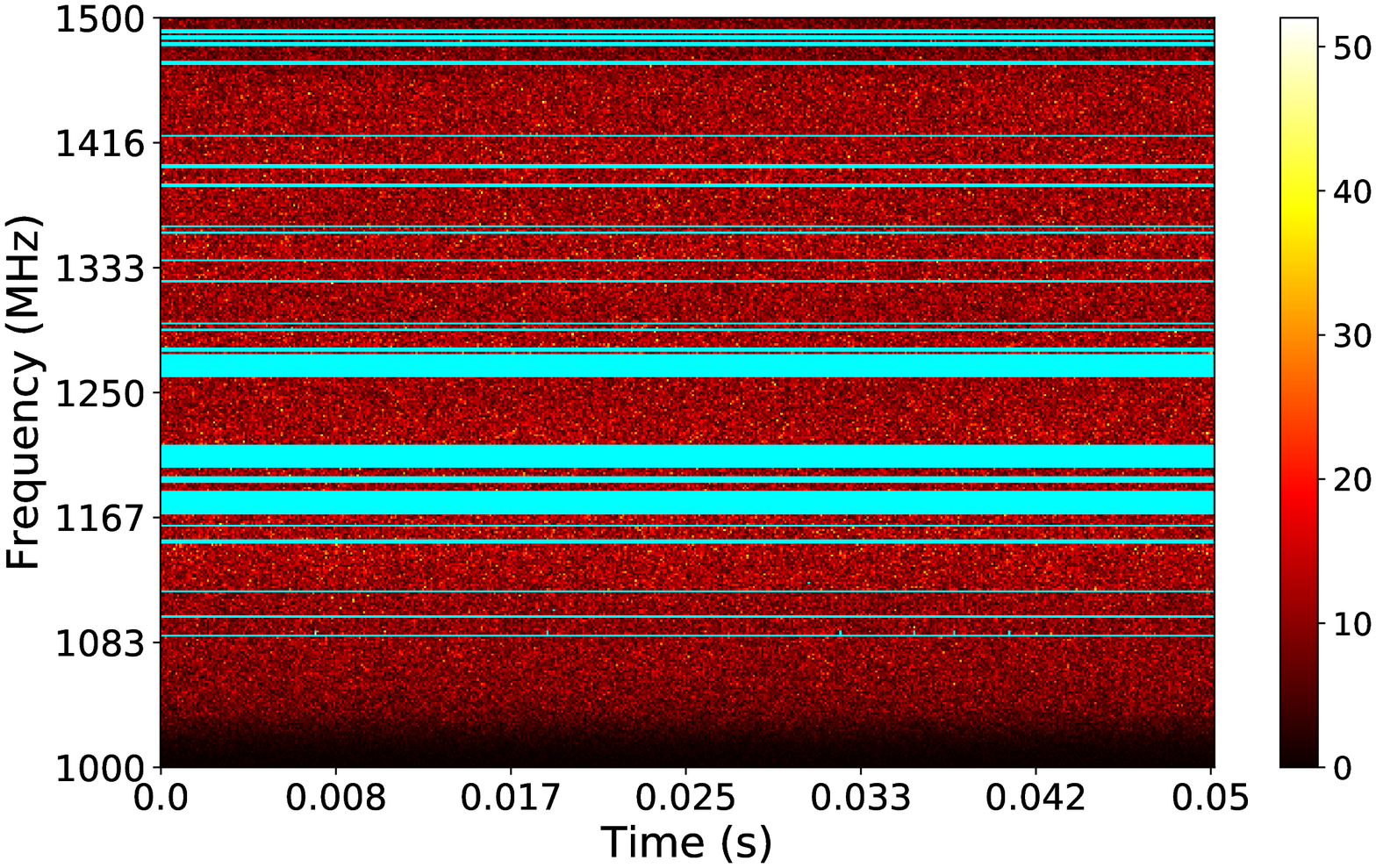}
	\label{fig:tf_full}
	}
	\subfloat[The mask manually flagged.]{
	\includegraphics[width=0.45\linewidth]{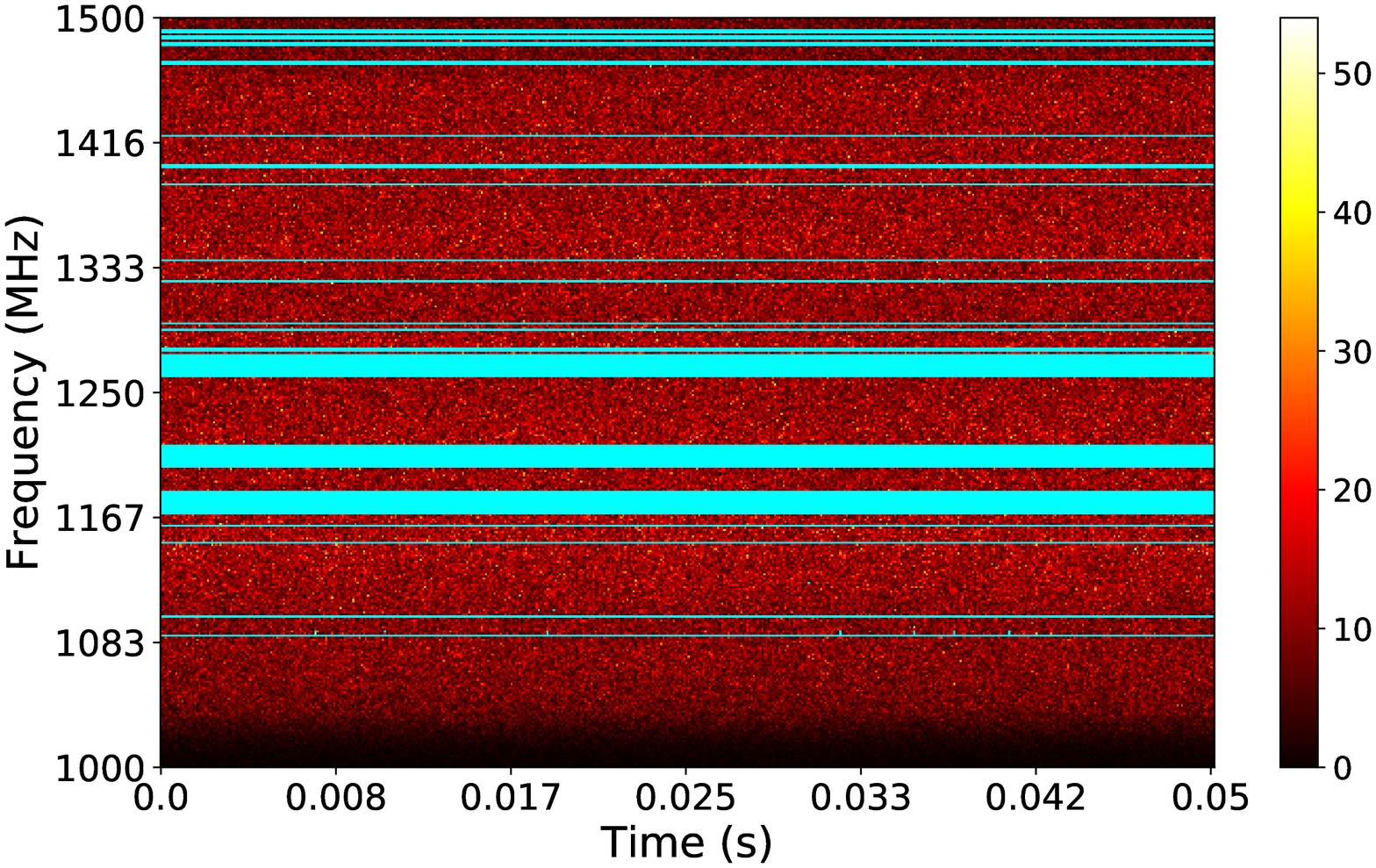}
	\label{fig:tf_hand}
	}
\caption{An example of FAST data for RFI detection and its labeled mask. (b) The computational results based on the procedures in section 3.2.2 for band RFI. (c) The result after flagging both band RFI and blob RFI using the ArPLS-ST method.  The result in (b) is blacker than the (c) since the blob RFI with higher intensity has not been removed in the former.}
\label{fig:label}
\end{figure*}

To detect the band RFI, the SumThreshold is performed on the SED curve after removing the estimated baseline. Experiments show that the protrusions above the horizontal line are detected excellently by the proposed scheme (Figure \ref{fig:global} b). These protrusions result from the energy differences between the RFI-contaminated pixels and the interference-free pixels. Therefore, the pixels on the frequency band corresponding with the detected protrusions on the SED curve will be flagged as band RFI (Figure \ref{fig:tf_band}).

Actually, there may exist some significant different intensities among the pixels contaminated by band RFI. The differences result from the variation in received power even though the telescope continuously received interference. This variation may come from several effects, such as intrinsic variation of the interference, change of propagation environment, and instrumental effects \citep{offringa2012algorithms}, etc. These differences in energy can result in leak detection for the RFI mitigation methods directly using the time-frequency image. Therefore, we remove the pixels corresponding to the flagged frequency channels on SED curves. However, this band RFI mitigation method has the potential possibility of bringing about some false positives. The false positives of this type occur between two strong band RFI in case of the uncontaminated pixels covering a small ratio of the area in the sub-integration being processed. This ratio should be so small that the corresponding frequency band can trigger the threshold in the SED curve. The probability of occurrence depends on the duration length of a sub-integration, and a short duration helps reduce this kind of possibility. Therefore, our experiments show that for FAST data this kind of negative possibility is insignificant and acceptable (Table \ref{tab:result}).

\subsubsection{Blob RFI detection}\label{sec:method:RFI:PontLikeRFI}

\begin{figure*}
    \centering
    \includegraphics[width=0.8\linewidth]{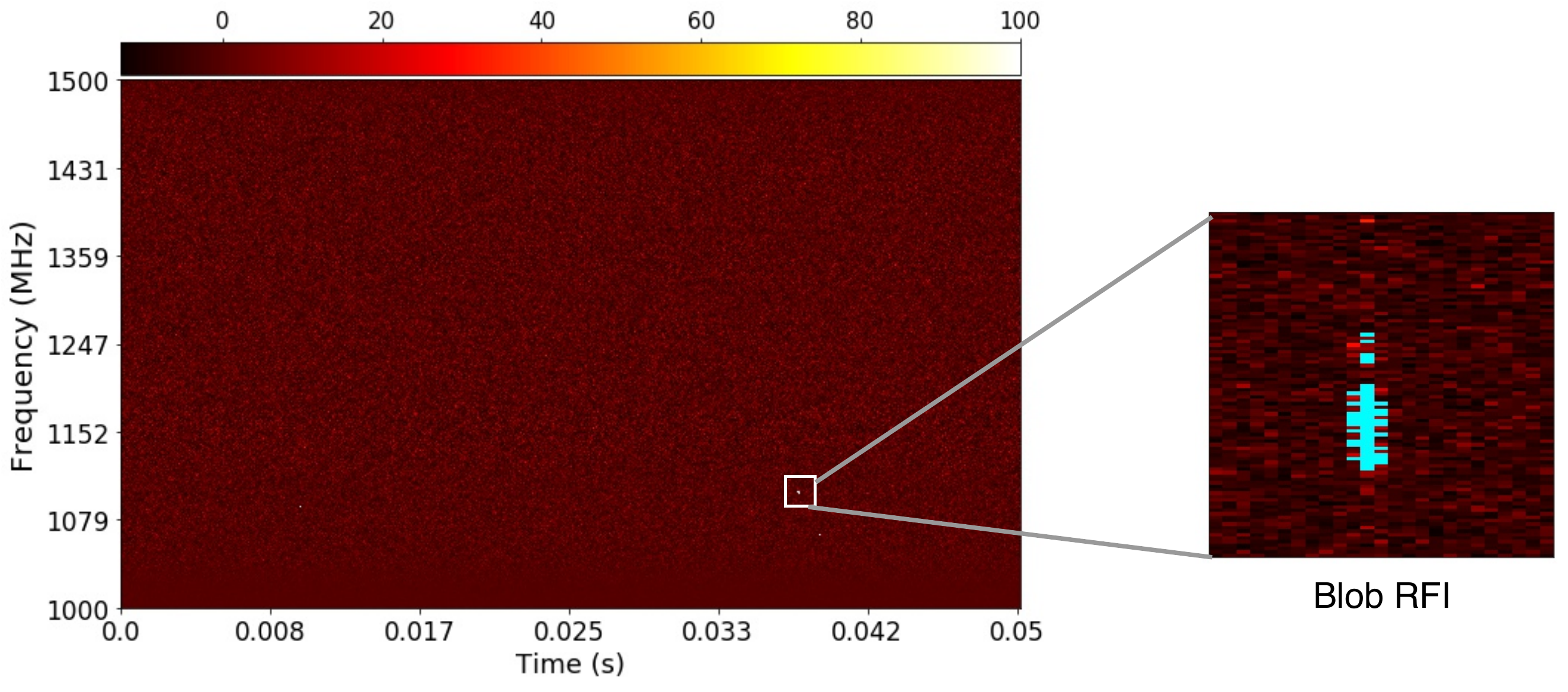}
    \caption{The result of blob RFI detection. }
    \label{fig:point_rfi}
\end{figure*}

After removing the estimated baseline from a time-frequency image (sections \ref{sec:method:baselinefitting:image}, \ref{sec:method:RFI:BandRFI}), we can get some results similar with the Figure \ref{fig:bl_tic} (b). After the band RFI is removed, the results are fed to the SumThreshold for blob RFI detection. Blob RFI bursts often exist with a certain duration both in the time and frequency direction. Therefore, the SumThreshold is executed iteratively and alternately along the time and frequency axis in a 2D image with the detection window increasing from 1 to a preset maximum width. The flagging procedure naturally starts from a large threshold for strong RFI and then, the threshold decreases exponentially. Finally, it outputs the mask indicating the position of the RFI (Figure \ref{fig:point_rfi}).

\section{Application to the FAST data and discussion}\label{sec:Results_Discussion}

To investigate the effectiveness of the proposed scheme, some quantitative evaluations and comparisons with several representative methods are conducted on real radio astronomy data (section \ref{sec:data}) for RFI detection. In this section, we first introduce experimental setting, and then present the experimental results.

\subsection{Experiment setting}
In this experiment, the proposed scheme is compared with five other methods: rfifind from PRESTO\footnote{\url{https://github.com/scottransom/presto}}, SEEK\footnote{\url{https://github.com/cosmo-ethz/seek}}'s Sumthreshold implementation with or without a morphological scale-invariant rank (SIR) operator\footnote{For convenience, this work uses SumThreshold (with SIR) and SumThreshold (without SIR) as the abbreviations for the cases of the traditional SumThreshold with a SIR operator and the traditional SumThreshold without a SIR operator, respectively.}, 1D polynomial fitting-SumThreshold (TPF-ST) and 1D Gaussian filter-SumThreshold (GF-ST). Each of the six RFI fagging methods was evaluated on 100 time-frequency images (section \ref{sec:data}). To make the evaluation results fair and not favor any automatic method, the ground-truth labels are generated by marking the RFI manually on the time-frequency images. It is worth noting that the methods are only tested for visually present RFI. The SIR operator is meant to detect RFI samples that are under the noise and invisible. Such samples would be counted as false positives. The performances of these methods are evaluated by Accuracy, False Positive Rate (FPR), False Negative Rate (FNR) and F1 score. Besides, the implementations of the last four methods are also based on SEEK (a python library), which makes the execution time of these methods comparable except for rfifind.

Among the five RFI detection methods, rfifind is the unique one that is not of a thresholding method. The operations for RFI detection in the TPF-ST and GF-ST are the same as those in the ArPLS-ST except for the baseline fitting methods. At the meantime, TPF and GF need to be executed alternatively with thresholding algorithms due to their sensitivity to the RFI. However, the ArPLS only needs to be executed once before detecting the RFI. These three methods all detect the band RFI on the SED curve and identify the blob RFI on the time-frequency image utilizing the SumThreshold methods. As for the traditional SumThreshold, detection of all types of the RFI is performed on the time-frequency image.

The parameters (e.g. thresholds in the SumThreshold, smoothness parameter $\lambda$ in ArPLS, etc.) that need to be determined are optimized by maximizing the F1 score. This is because the F1 score is capable of measuring the overall performance of the methods when facing with the classification on the imbalanced data. The imbalance refers to the situation that the RFI-free pixels are much more than RFI-contaminated pixels. The optimization of the parameters is implemented through a grid search.


\subsection{Experimental results and discussion}

The performance metrics of the RFI detection on the FAST data are presented in Table \ref{tab:result}. On the whole, ArPLS-ST outperforms the other methods, especially on the accuracy, FNR, F1, and efficiency. The performances of the last three methods (TPF-ST, GF-ST, ArPLS-ST) are better than the traditional SumThreshold\footnote{The traditional SumThreshold refers to the SumThreshold (with SIR) and the SumThreshold (without SIR) introduced by \citet{offringa2010post}.}. The main difference between the implementations of these three methods, and the traditional SumThreshold is that the traditional SumThreshold fits the baseline and detects the band RFI in a 2D time-frequency image, while the other methods do these on the integration curve SED. The experimental results in Table \ref{tab:result} show that the methods \{TPF-ST, GF-ST, ArPLS-ST\} achieve much better performance than the traditional SumThreshold, and indicate the superiority of baseline estimation on SED curve.

On the other hand, the intensities of band RFI are much weaker than that of the blob RFI. Therefore, the thresholds for band RFI performed on SED curves are set smaller than those for blob RFI on the time-frequency images in TPF-ST, GF-ST, and ArPLS-ST. In traditional SumThreshold, there is just one threshold for all types of RFI. To detect the band RFI as much as possible, a small threshold should be chosen. However, this small threshold is likely to result that too many non-contaminated pixels with slightly high intensity are mistakenly detected as blob RFI. On the contrary, a large threshold can bring about leak detections around a detected band RFI, and a high FNR. Therefore, the last three rows in Table \ref{tab:result} show that designing different thresholds for different types of RFI is essential to substantially improve the RFI detection performance.

In addition, the traditional SumThreshold algorithm applies a SIR operator
\citep{offringa2012morphological, van2016efficient} to enlarge the flag mask and avoid the failure detection for the RFI with weak intensities in the presence of the variation in received power \citep{offringa2012morphological}. After applying SIR operator, the performance, based on FNR and F1, of the traditional SumThreshold is improved. However, it raises the FPR from 2.05\% to 7.98\%. Therefore, this work utilizes the thresholding method on a SED curve and remove all the pixels corresponding to the flagged frequency channels to handle the variation of the received power. This optimization dramatically improves the performance of the RFI mitigation method without bringing about too many false positives (the last row in Table \ref{tab:result}). This work also tried the SIR operator for blob RFI detection but it did not get the result we expected. Although the main parts of the blob RFI are stronger than other signals (such as pulsar signals, band RFI, background information, etc.), its wings are weak and presumably continue under the noise (Figure \ref{fig:point_rfi}). The detection of this kind weak RFI pixels can be improved by SIR operator. However, some of the similar weak pixels also can be non-RFI pixels which result in false positives by SIR operator. At the same time, the SIR operator needs more computation and decreases efficiency. Therefore, the proposed scheme utilizes the SumThreshold without SIR for blob RFI detection.

\begin{figure*}
\centering
	\subfloat[The detection result of rfifind.]{
	\includegraphics[width=0.45\linewidth]{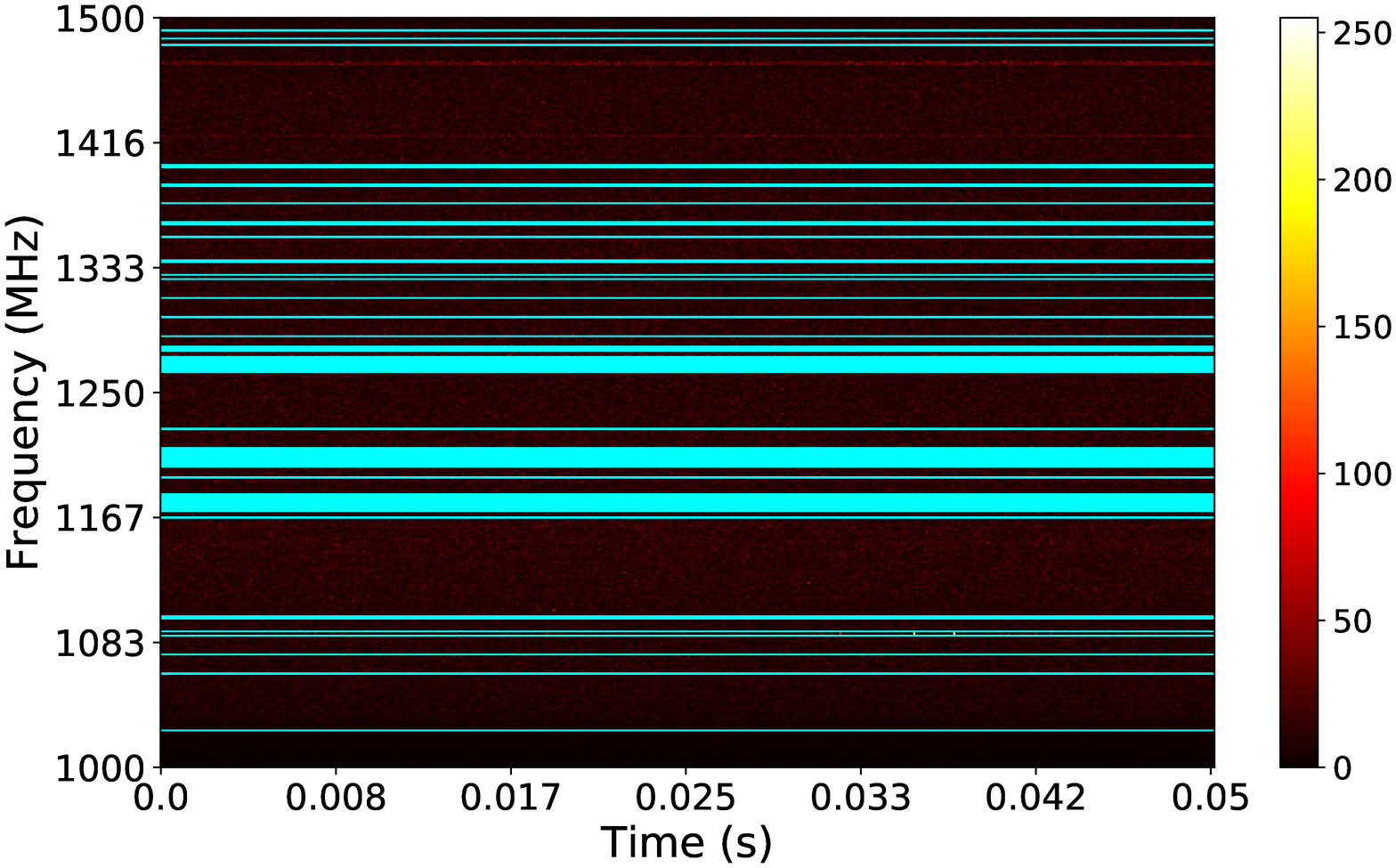}
	\label{fig:rfifind}
	}
	\subfloat[The detection result of the SumThreshold (with SIR).]{
	\includegraphics[width=0.45\linewidth]{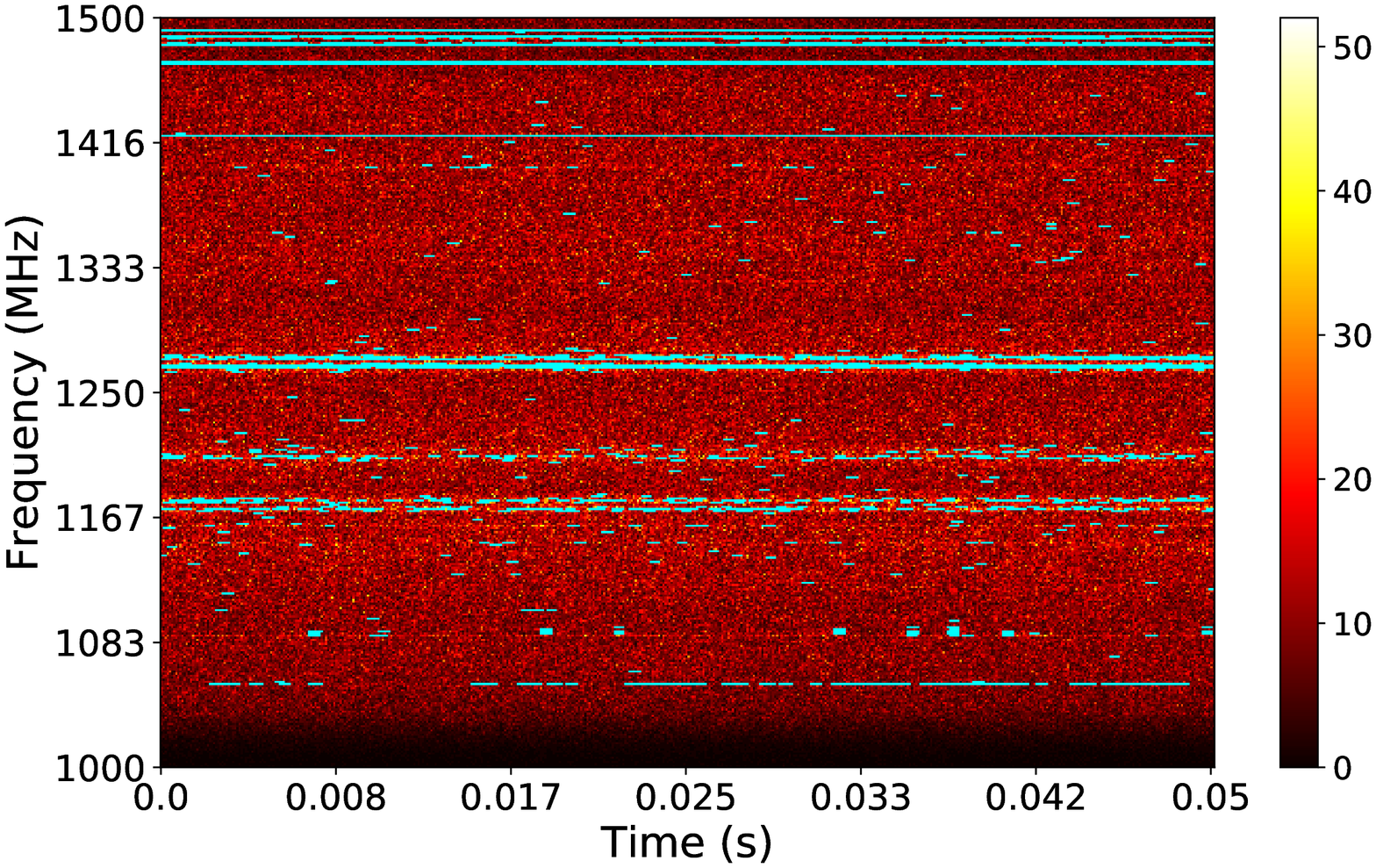}
	\label{fig:stsir}
	}
	\vfill
	\subfloat[The detection result of the SumThreshold (without SIR)]{
	\includegraphics[width=0.45\linewidth]{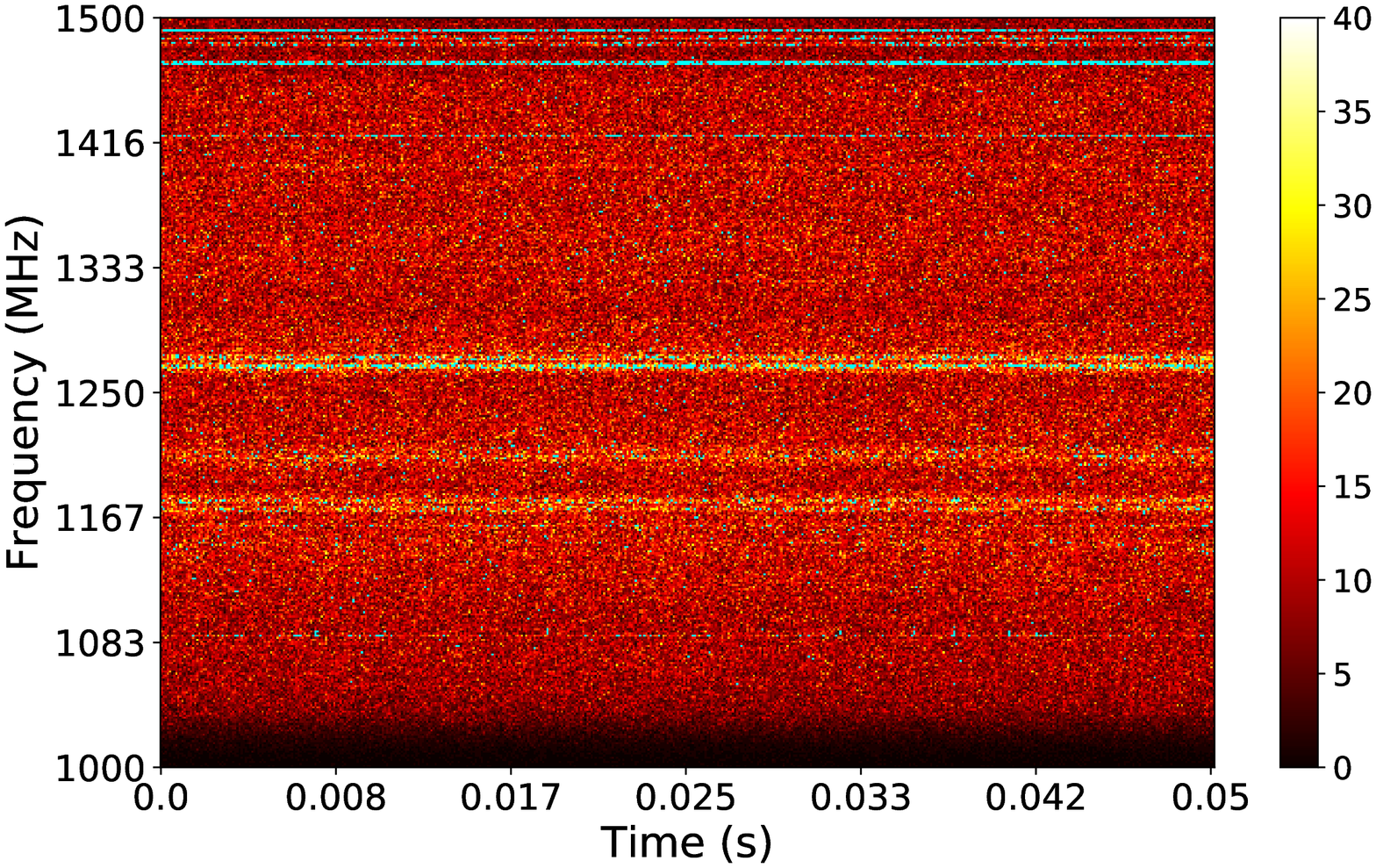}
	\label{fig:st}
	}
	\subfloat[The detection result of PF-ST.]{
	\includegraphics[width=0.45\linewidth]{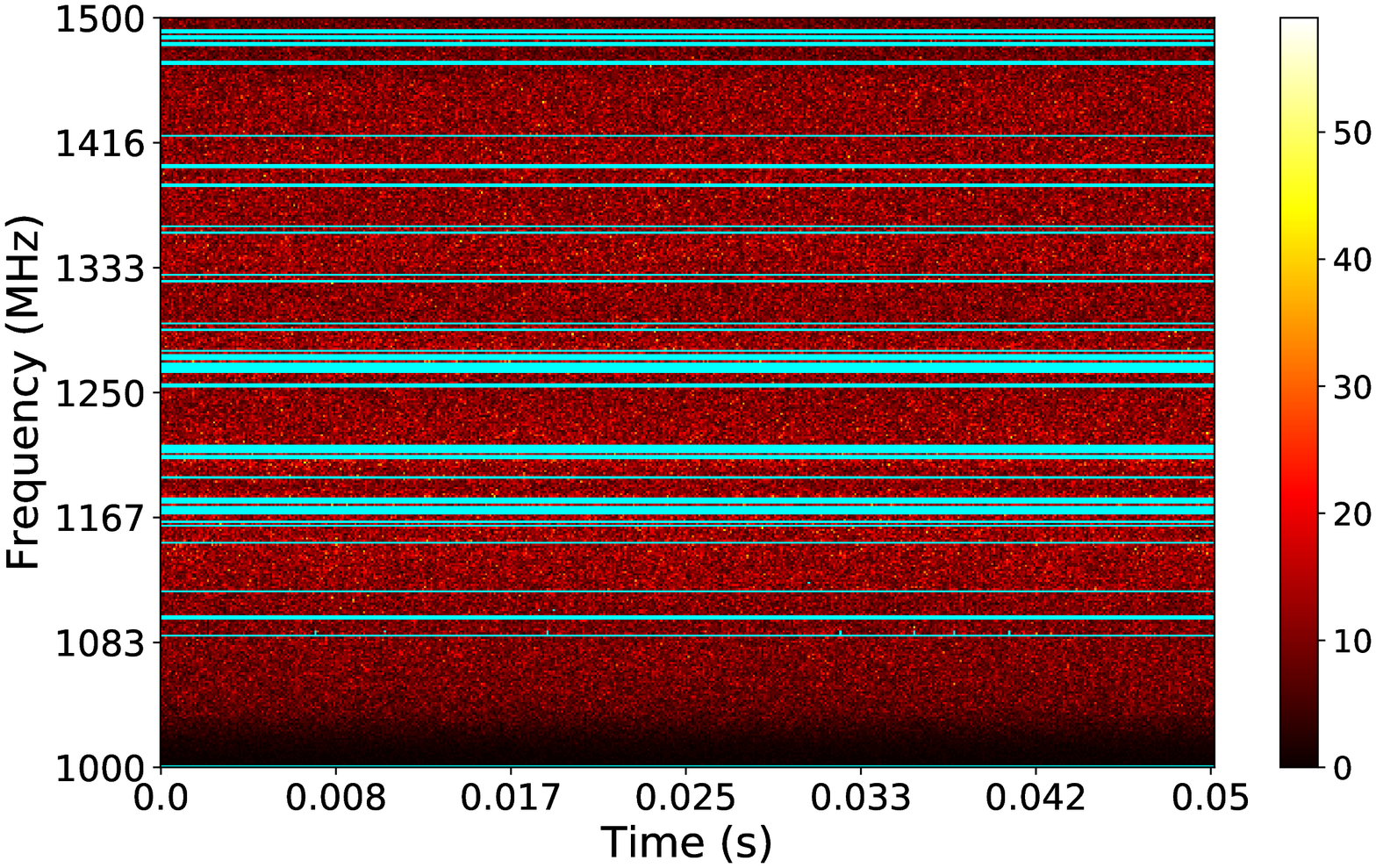}
	\label{fig:pf_st}
	}
	\vfill
	\subfloat[The detection result of GF-ST.]{
	\includegraphics[width=0.45\linewidth]{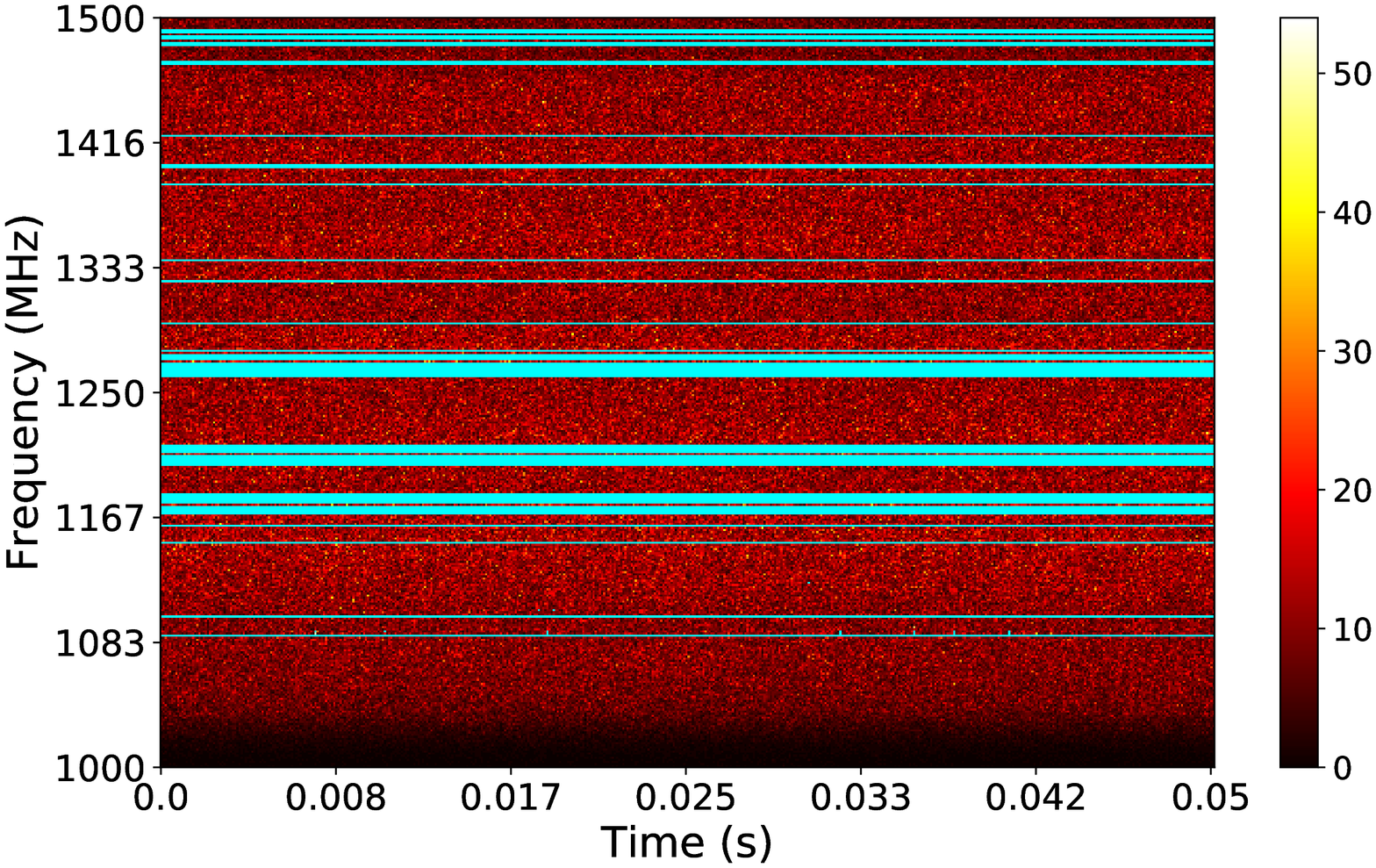}
	\label{fig:pf_st}
	}
	\subfloat[The detection result of ArPLS-ST.]{
	\includegraphics[width=0.45\linewidth]{figures/G69_04+0_00_tracking-M14_0048_fits_pls_result.eps}
	\label{fig:arpls_st}
	}
\caption{The detection result of six methods. The first subfigure looks black outside the RFI regions because the strong blob RFI has not been removed by rfifind and the intensity of the rest part is relatively weak.}
\label{fig:result}
\end{figure*}

As for the TPF-ST and GF-ST, the difference between these two methods and the proposed ArPLS-ST is on the baseline fitting procedure. The experiments in the last three rows of Table \ref{tab:result} show that the proposed ArPLS method can obtain a more appropriate estimation for baseline than the tile-based polynomial fitting and Gaussian filter method. Although the proposed scheme has a slightly higher FPR than the GF-ST, the false positives in the ArPLS-ST are always the ``small burr'' in the integration. On the contrary, the GF-ST always makes some mistakes in the multiple-peaks regions, which is likely to have some more severely adverse impacts on subsequent analysis and application. As shown in Figure \ref{fig:result}, the TPF-ST also suffers from the multi-peaks problem on frequency ranges 1220-1244 MHz, and 1281-1305 MHz, etc.

The first method, rfifind, is totally different from the other methods described above. It mainly detects the broadband RFI with a short duration and strong narrow-band RFI \citep{ransom2001new}. The broadband RFI with a short duration is detected by performing a time-domain clipping of the curve integration by channels and the strong narrow-band RFI is detected based on the computational result of Fast Fourier Transform Algorithm. However, the rfifind is not able to detect relatively weak RFI and blob RFI. Therefore, the results in Table \ref{tab:result} show that the rfifind does not perform as well as the ArPLS-ST overall.

Figure \ref{fig:result} shows the results of the six RFI flagging methods on one time-frequency image. It is found that the results of the last three methods (TPF-ST, GF-ST, ArPLS-ST) are similar in general. Besides, the rfifind is unable to flag the blob RFI. Therefore, the area outside the band RFI regions is dark due to the high intensity of blob RFI (Figure \ref{fig:rfifind}). As for the traditional SumThreshold, it can not flag the band RFI completely, especially in the absence of the SIR operator.
%

\begin{table*}
	\centering
	\caption{Results of six RFI flagging methods evaluated on 100 time-frequency images (section \ref{sec:data}). Each of the methods TPF-ST, GF-ST, ArPLS-ST, rfifind, SumThreshold (with SIR) and SumThreshold (without SIR) refers to a full pipeline of detecting both band RFI and blob RFI. ``Execution time'' consists of the computation time of baseline fitting and RFI detection on one 4096 $\times$ 1024 image randomly selected from the 100 time-frequency images (section \ref{sec:data}), and the Accuracy, FPR, FNR and F1 are computed from all of the 100 images.}
	\label{tab:result}
	\begin{tabular}{llllll} 
\hline
Method       & Accuracy & FPR  & FNR   & F1    & Execution time\\
\hline
Rfifind      & 88.54    & 3.14 & 55.60 & 58.22 &     \textit{not comparable}\\
SumThreshold(with SIR) & 82.98    & 7.98 & 64.95 & 48.96 &                16900$\pm$ 66 ms\\
SumThreshold(without SIR) & 83.80    & 2.05 & 91.25 &  16.08&                16500$\pm$ 17 ms\\
TPF-ST        & 93.51    & 3.20 & 23.89 & 82.63 &  705$\pm$9.35ms              \\
GF-ST        & 96.60    & $\mathbf{1.08}$ & 16.08 & 87.39 &752$\pm$17.5ms                \\
ArPLS-ST     & $\mathbf{97.95}$    & 1.53 & $\mathbf{4.78}$  & $\mathbf{93.65}$ &  $\mathbf{534 \pm4.5ms}$ \\
\hline
	\end{tabular}
\end{table*}


Logically, every method in the rfifind, SumThreshold (with SIR), SumThreshold (without SIR), ArPLS-ST, TPF-ST and GF-ST consists of two procedures: baseline fitting and RFI detection (band RFI and blob RFI). To fit the baseline, the traditional SumThreshold utilizes Gaussian filter \citep{offringa2010post} in a 2D time-frequency image; the last three methods (TPF-ST, GF-ST and ArPLS-ST) conduct computations on a 1D SED curve, which result in a more efficient implementation than the traditional SumThreshold. In the RFI detection, the five methods cost similar running time. Therefore, the TPF-ST, GF-ST and ArPLS-ST are much more efficient than the SumThreshold (with SIR) and SumThreshold (without SIR) (Table \ref{tab:result}).

It is worth noting that all the listed methods except for the rfifind are implemented on Python without any optimization. However, the execution speed of Python code is significantly lower than that of Fortran, C, or C++ because the Python is an interpreted language, not compiled, and its efficiency is affected by Global Interpreter Lock (GIL). Therefore, although the comparison on execution time between them is relatively fair in Table \ref{tab:result}, the efficiency of the TPF-ST, GF-ST, SumThreshold (with or without SIR) and ArPLS-ST can be increased if they are implemented using C, C++ or Fortran, and optimized (such as parallel computing, GPU acceleration).

In summary, the scheme ArPLS-ST is proposed for radio data processing. Experiments on the FAST data show that this scheme can effectively detect the RFI. It provides a fast and accurate baseline estimation method based on the SED curve to reduce the potentially negative influences from the inconsistency of the receiver response, accurately locate the RFI regions. Several identification strategies are designed for detecting RFI.

In future, some potential improvements and extensions still can be made:
\begin{itemize}
    \item \textbf{Parameter setup}. There are two types of parameters that need to be set in the ArPLS-ST. One is the smoothness parameter $\lambda$ in the ArPLS algorithm. This work experientially set it to 10000, a constant, which could obtain satisfactory results for all of the available FAST data. However, it may not be good enough to handle the complex radio environment in other situations. In practice, the smoothness parameter can be automatically determined by some statistics which quantify the characteristics of original integration curve. Another one is the threshold in the SumThreshold algorithm. As mentioned in Section \ref{sec:method}, the threshold is determined by the K$\sigma$ criterion which concentrates on the aggregation of the pixel intensity distribution. In fact, it may be a more natural and robust way to set this kind of parameter through the pixels far away from the cluster. Some outlier detection techniques may be taken into account to solve this problem in future.
    \item \textbf{More accurate band RFI flagging strategy}. The band RFI flagging strategy in the ArPLS-ST will remove all of the pixels within the marked channel in one sub-integration. This trigger-remove-all scheme may potentially result in some false positives to some extent. An accurate band RFI flagging strategy that has the ability to identify the band RFI with different durations may be a better choice.
    \item \textbf{Distinguishness between the signal of interest and RFI}. The key of the threshold-based RFI flagging methods is that the energy of the RFI bursts is much stronger than that of non-RFI data and signal of interest. The traditional thresholding algorithms will identify the strong signal as the RFI. This kind of false positive causes huge losses for the research and is not allowed to happen in practice. Therefore, distinguishing them according to their characteristics is the most important and urgent task for the thresholding-based RFI flagging methods. We designed a novel method to distinguish between the signals of interest and the detected candidate RFI.
    \item \textbf{Software availability}. The Python software package of this work will be updated whenever possible at \url{http://zmtt.bao.ac.cn/GPPS/RFI} for open usage, given the proper citation to this paper.
\end{itemize}

 \section*{Acknowledgements}

X. L. and Q. Z. were supported by the National Natural Science Foundation of China
(Grant Nos. 61075033), the Natural Science Foundation
of Guangdong Province (No. 2020A1515010710). 
C.W. was supported by the National Natural Science
Foundation of China (U1731120).


\section*{Data availability}

A Python software package of this work and sample data are available at \url{http://zmtt.bao.ac.cn/GPPS/RFI}



\bibliographystyle{mnras}
\bibliography{example} 



%
%
%


\bsp	
\label{lastpage}
\end{document}